# A Shock-Optimized SECE Integrated Circuit

Adrien Morel, *Member, IEEE*, Anthony Quelen, Pierre Gasnier, Romain Grézaud, Stéphane Monfray, Adrien Badel, Gaël Pillonnet, *Senior Member, IEEE*.

*Abstract*— This paper presents a fully integrated, self-starting shock-optimized Synchronous Electric Charge Extraction (SECE) interface for piezoelectric harvesters (PEHs). After introducing a model of the electromechanical system under shocks, we prove that the SECE is the most appropriate electrical interface to maximize the harvested energy from our PEH. The proposed interface is then presented, both at system- and transistor-levels. Thanks to a dedicated sequencing, its quiescent current is as low as 30nA. This makes the proposed interface efficient even under time-spaced shocks occurring at sporadic and unpredictable rates. The circuit is for instance able to maintain its self-powered operation while harvesting very small shocks of $8\mu J$ happening every 100 seconds. Our chip was fabricated in CMOS 40nm technology, and occupies a 0.55mm$^2$ core area. The measured maximum electrical efficiency under shocks reaches 91%. Under shocks, the harvested energy by the proposed shock-optimized SECE interface is 4.2 times higher than using a standard energy harvesting circuit, leading to the best shock FoM among prior art.

## I. Introduction

In the last two decades, energy harvesting has been widely investigated as a potential alternative or complement to batteries in order to power sensor nodes. Indeed, in many applications, the use of battery is not the most appropriate choice due to size and thermic constraints [1], or when their replacement is either costly or hazardous [2]. Energy harvesting consists in scavenging the available energy in the environment and to turn it into storable usable electrical energy. Ambient energy can take many forms: solar [3], thermal [4], chemical [5], and mechanical [6]. In closed confined environments, where there are low thermal gradients nor solar radiations, mechanical vibrations may be the most important energetic source. Indeed, vibrations harvesters energy densities under real conditions usually go from 4 to $800\mu W.cm^{-3}$ [7].

In order to convert mechanical energy into an electrical form, three main transduction mechanisms have been investigated in the literature: electrostatic, electrodynamic, and piezoelectricity [8]. Among those, piezoelectric energy harvesters (PEH) combine relatively high power density [9] while being scalable for sub-cm scale PEH [10].

Under any excitation, PEHs produce AC voltage that cannot be directly stored in a capacitance or delivered to a sensor node. Hence, the main aim of the electrical interface is to rectify this AC voltage. The first interface that has been proposed in prior art is the Standard Energy Harvesting (SEH) interface. It consists in a full bridge rectifying the piezoelectric voltage and directly connected to a storage capacitance. This interface is not optimal as it only extracts a small part of the energy stored in the harvester [11].

In order to increase the harvested energy under periodic excitations, non-linear interfaces such as Synchronous Electrical Charge Extraction (SECE) [12], Energy Investing [13], or Synchronized Switch Harvesting on Inductor (SSHI), also called Bias-flip by the integrated circuit community [14] have been developed. However, in most of today's applications, vibrations are not periodic and mechanical stimuli occur at unpredictable rates [15]. SSHI interfaces naturally seemed to be the most appropriate candidate for harvesting shocks as they exhibit outstanding performances under periodic excitations [16]. However, the SSHI strategy presents inherent weaknesses while harvesting shocks, since SSHI strategy efficiency is highly dependent on the voltage across the storage element, and thus requires an impedance matching block [17]. This impedance matching is challenging, due to the sporadic nature of shocks (variable amplitude), making them quite unpredictable.

In this paper, we propose an electrical interface based on the SECE strategy which has been optimized to harvest shocks vibrations. First, we develop the model of the electromechanical harvester, based on an energy balance modelling. By numerical simulations, we then use this model to compare different electrical interfaces efficiencies, i.e. SEH, SSHI and SECE interfaces. From these simulations, we conclude that the most appropriate strategy is, using our selected PEH, the SECE. After detailing the SECE strategy, we propose a theoretical model of its efficiency under shocks, which accurately predicts the electrical harvested energy from the harvester. Finally, we present our self-starting, battery-less, integrated energy harvesting interface based on the SECE strategy which has been optimized to work under shock stimulus, both at system and transistor level. Due to the sporadic nature of mechanical shocks which implies long periods of inactivity and energy harvesting periods, the

This research was, in part, funded by the French Inter-Ministerial Fund (FUI), through HEATec project, and by STMicroelectronics. A. Morel, A. Quelen, P. Gasnier, R. Grézaud and G. Pillonnet are with Univ. Grenoble Alpes, CEA, LETI, MINATEC, F-38000 Grenoble, France (e-mail: adrien.morel@cea.fr, gael.pillonnet@cea.fr). S. Monfray is with STMicroelectronics, Grenoble, France. A. Morel and A. Badel are with Univ. Savoie Mont Blanc, SYMME, F-74000, Annecy, France (e-mail: adrien.badel@univ-smb.fr).

interface's average consumption is optimized by minimizing the quiescent power between two shocks. A dedicated energy saving sequencing has thus been designed, reducing the static current. Our SECE-based circuit features shock and periodic vibrations harvesting capabilities. It has been experimentally validated and compared to previously reported interfaces.

## II. MODELLING OF THE HARVESTER UNDER PULSED EXCITATION

### A. Piezoelectric harvester modelling

A piezoelectric energy harvester (PEH) is constituted of a piezoelectric ceramic bonded on a mechanical resonator, as shown in Fig.1.a. Under an external vibration, the resonator starts oscillating, applying a strain on the piezoelectric material. Due to the direct piezoelectric effect, electric charges are generated in the material. Then, those charges can be collected thanks to the electrical interface. If we assume linearity of both the mechanical oscillator and the piezoelectric material, such linear harvester can be modeled by (1) [18].

$$\begin{cases} M\ddot{x} + D\dot{x} + K_{sc}x + \alpha v_p = F \\ i_p = \alpha \dot{x} - C_p \dot{v}_p \end{cases} \quad (1)$$

Where $F$, $x$, $i_p$ and $v_p$ stand for the input vibrational force, the displacement of the tip mass, the current extracted in the interface circuit, and the voltage across the electrodes of the piezoelectric material, respectively. $M$, $D$, $K_{sc}$, $C_p$ and $\alpha$ stand for the equivalent dynamic mass of the system, its mechanical damping, its short-circuit stiffness, the piezoelectric material clamped capacitance, and the piezoelectric coefficient (force factor). The model (1) is relatively accurate as long as the mechanical oscillator linearity hypothesis is respected, and the excitation's frequency remains close to the short-circuit resonant frequency of the oscillator [18]. Furthermore, this model assumes that the higher mechanical mode will not be excited and will not contribute much to the mechanical displacement.

In order to model such systems under pulsed regime, we will consider that every shock can be associated with a certain potential energy quantity $E_{in}$ transferred instantaneously to the mechanical oscillator. Fig.1.b shows a standard piezoelectric harvester and its associated electrical model.

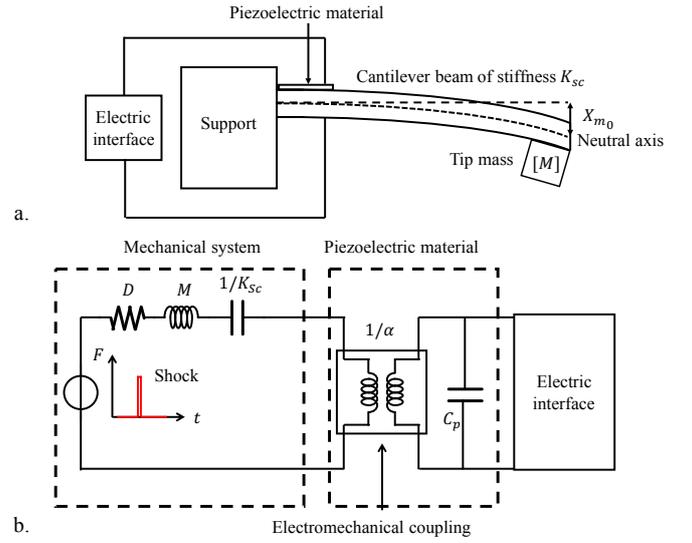

Fig.1. a) Piezoelectric harvester representation and b) its associated electrical model

Throughout this paper, simulation results and experimental results are derived based on a MIDE piezoelectric generator (PPA1011) whose characteristics have been determined thanks to an impedance analyzer. Those characteristics are summarized in Table I. Two normalized parameters commonly used in the literature [18,19] are introduced:

- The squared electromechanical coupling coefficient, $k_m^2 = \alpha^2 K_{sc}^{-1} C_p^{-1}$.
- The mechanical quality factor of the harvester, $Q_m = \sqrt{MK_{sc}}/D$.

TABLE I
PIEZOELECTRIC HARVESTER CHARACTERISTICS

| PEH characteristics | $M$ | $K_{sc}$ | $D$ | $\alpha$ | $C_p$ | $k_m^2$ | $Q_m$ |
|---|---|---|---|---|---|---|---|
| Values | 5.67 | 1275 | 29.3 | $6.3e^{-04}$ | 43 | 0.8% | 92 |
| Units | $g$ | $N.m^{-1}$ | $g.s^{-1}$ | $N.V^{-1}$ | $nF$ | – | – |

### B. Energy balance considerations

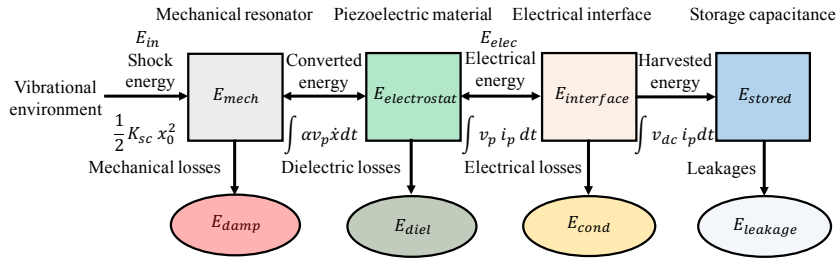

Fig.2. Energy balance modelling of the electromechanical system

The energy flow associated with a standard piezoelectric harvester is shown in Fig.2. $E_{mech}, E_{electrostat}, E_{interface}$, and $E_{stored}$ are the potential energies contained in the mechanical resonator, the piezoelectric material, the electrical interface and the storage element, respectively. $E_{damp}, E_{diel}, E_{cond}$, and $E_{leakage}$ are the energy losses due to mechanical friction, dielectric losses, resistive electrical paths, and storage leakage, respectively. We assume in the following that the dielectric losses $E_{diel}$ inside the piezoelectric material are negligible compared to the mechanical and electrical ones. The electrical losses in the interface circuit $E_{cond}$ are circuit and topology dependent. In order to evaluate the interface potentials without any electronic-dependent efficiency considerations, we define $E_{elec}$ as the extracted energy from the piezoelectric material, before it goes through the electrical interface.

Multiplying (1) by the derivative of the displacement, $\dot{x}$, and integrating it between $t_0$ (time associated with a shock occurrence) and $t$, yields the energy balance equation of the electromechanical system given by (2).

$$E_{mech} + E_{electrostat} + E_{elec} + E_{damp} = E_{in} \quad (2)$$

Expressions of those energies are given by system (3).

$$\begin{cases} E_{kin} = \frac{1}{2} M \, \dot{x}^2(t) \\ E_{pot} = \frac{1}{2} K_{sc} \, x^2(t) \\ E_{mech} = E_{kin} + E_{potential} \\ E_{damp} = D \int_{t_0}^{t} \dot{x}^2(t) \, dt \\ E_{electrostat} = 1/2 \, C_p \, v_p^2(t) \\ E_{elec} = \int_{t_0}^{t} v_p(t) i_p(t) \, dt \end{cases} \quad (3)$$

Where $E_{kin}$ and $E_{pot}$ represent the kinetic energy and potential energies in the mechanical resonator, respectively. Obviously, the goal of the harvesting interface is to maximize the harvested energy. This is done by:

- Finding the best harvesting strategy to maximize the extracted electrical energy $E_{elec}$. Contrarily to the periodic case, the extracted electrical energy should always be maximized and not be equal to $E_{damp}$. Thus, we should maximize the mechanical-to-electrical energy conversion, associated with $\eta_{electromech}$:

$$\eta_{electromech} = \frac{E_{elec}}{E_{in}} \quad (4)$$

- Minimizing the losses in the electronic interface. This will be done by choosing an adapted circuit topology and control circuit sequencing, as extensively explain in Section IV. Thus, we should maximize as well the electrical-to-electrical energy conversion, associated with $\eta_{elec}$.

$$\eta_{elec} = \frac{E_{stored}}{E_{elec}} \quad (5)$$

C. *Nonlinear electrical interfaces under shocks*

In order to harvest energy from piezoelectric harvesters, three main families of interface have been developed in prior art [20].

Standard Energy Harvesting (SEH) interfaces consist in directly connecting a rectifier followed by a storage element to the piezoelectric element. This interface does not require any control circuit nor active component. However, the harvested power with SEH interface is limited, as it only allows to convert a small part of the mechanical energy into electrical energy. Furthermore, its efficiency is highly dependent on the voltage across the storage element [21].

Synchronized Switched Harvesting on Inductor (SSHI) interfaces, also called bias-flip by the integrated circuit community, have been developed in order to enhance the harvested energy from piezoelectric harvesters. By inversing the voltage polarity in the piezoelectric element at the right instants thanks to an inductive switch, it is able to minimize the time where the energy is not extracted, and to maximize the accumulated charges in the material. This strategy has been widely investigated in prior art for maximizing the harvested energy from periodic excitations [22]. However, its efficiency is also very dependent on the voltage across the storage element [14].

Synchronous Electric Charges Extraction (SECE) interfaces consist as well in connecting an inductance to the piezoelectric material when the mechanical displacement reaches an extremum. However, contrary to SSHI interfaces, all the charges extracted from the piezoelectric material are not reinjected in the harvester, but directly stored in a storage element. This strategy is, under period excitations, theoretically up to four times more efficient than SEH [11]. Its main advantage is that its efficiency is independent on the voltage across the storage element.

Each interface has been simulated associated with the linear electromechanical model given by (1) implemented on Matlab@Simulink. The DC voltage across the storage element $v_{dc}$ has been fixed to $2.5V$ which is a standard voltage to supply sensor nodes, while the shock energy $E_{in}$ has been fixed to $57\mu J$. The model parameters correspond to the piezoelectric harvester used in the experimental part of this paper and have been previously given in Table I. The different energies introduced by (3) have been computed and are shown as a function of the time starting from the shock occurrence at $t_0 = 0$. The electrostatic energy $E_{electrostat}$, which appears to be really small compared to other energy forms, has not been represented in the results shown in Fig.3-5. Indeed, $E_{interface}$, which corresponds to the cumulative sum of all the previous electrostatic energy extremum in the case of the SECE, becomes quickly more important than $E_{electrostat}$ after a few vibrations periods.

As shown in Fig.3, the efficiency of the SEH interface is limited: most of the shock energy is dissipated as heat due to mechanical friction. The harvested energy can be improved using SSHI, as shown in Fig.4, which drastically reduces the losses due to mechanical friction. This is mainly due to the SSHI circuit effect, which rapidly converts the mechanical energy into electrical energy and damps the mechanical displacement, leading to low mechanical losses. SSHI interface efficiency depends on the voltage across the storage element. If this voltage is too low, almost no energy is harvested every semi-period of vibration. In another hand, SSHI efficiency could be even higher if the voltage across the storage element were more important. However, to realize such impedance matching under pulsed operation, we would have to know the shock energy at the instant the shock happens. Furthermore, the voltage across the storage element, should be ideally variable with time in order to maximize the energy extraction, which seems hardly realizable. It can be also noted that SSHI interfaces introduce important electrical losses that are linked with the charge inversion quality factor. Those losses can be limited with a precise and thorough circuit design [14], however it requires particular attention and constitute an additional design constraint. Finally, the SECE interface performances are shown in Fig.5. SECE interface greatly improves the harvested energy compared to SEH, and is even slightly more efficient than the SSHI interface. It may be noted that Fig.3-5 are only valid for the specific shock energy $E_{in} = 57\mu J$. If this shock energy were more important, the SEH and SSHI harvesting periods would be longer. In another hand, if the shock energy were lower, the SEH and SSHI harvesting periods would be shorter.

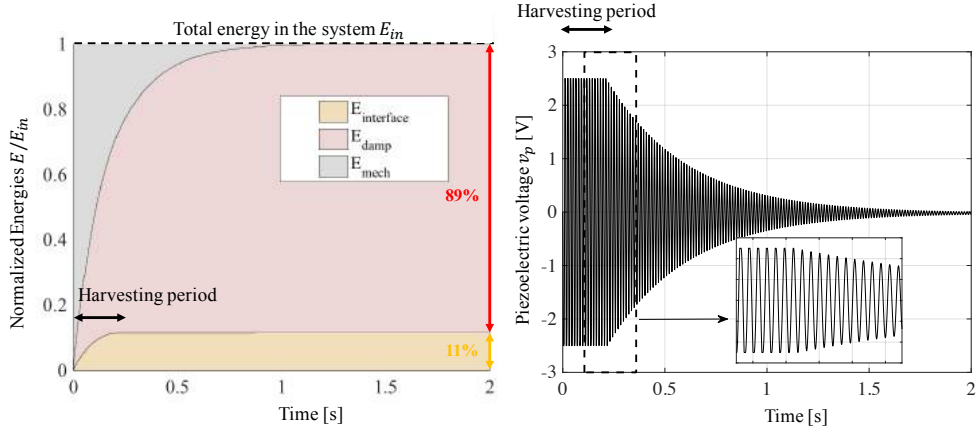

Fig.3. Normalized energies repartition with SEH interface

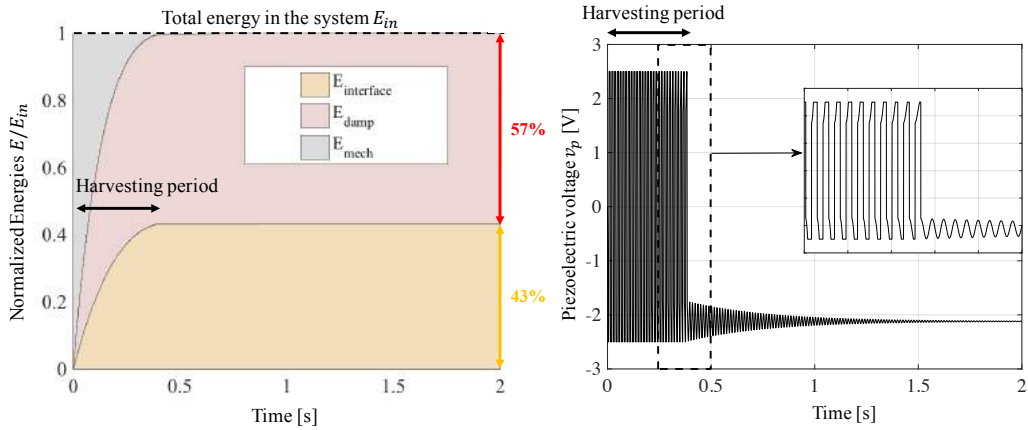

Fig.4. Normalized energies repartition with SSHI interface

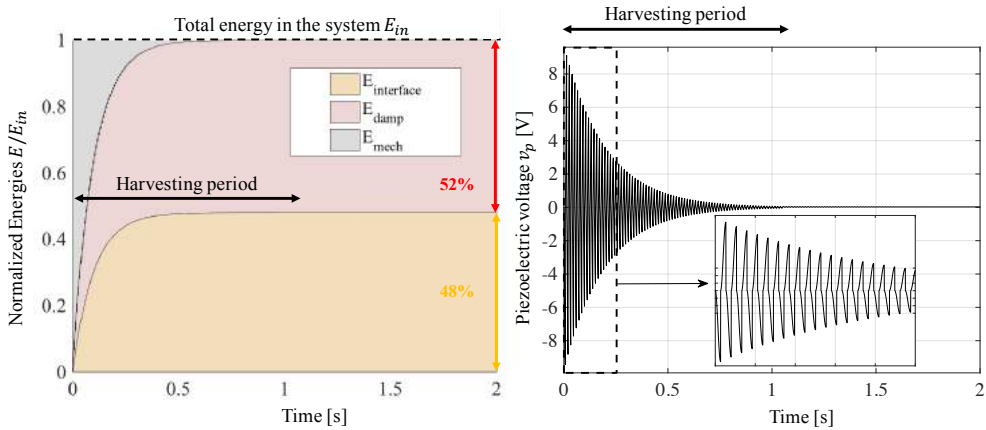

Fig.5. Normalized energies repartition with SECE interface

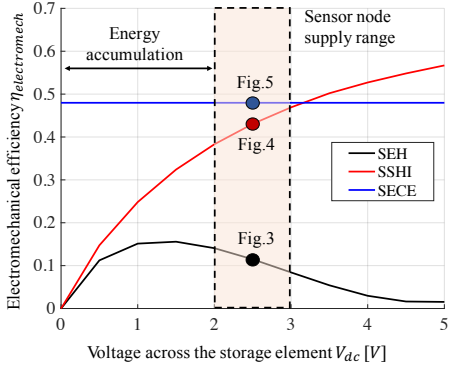

Fig.6. Simulated strategies efficiency as a function of the voltage across the storage element

Furthermore, as shown on Fig.6, making SSHI and SEH interfaces efficient usually requires the use of a DC/DC converter realizing an impedance matching. Because of the unpredictable and sporadic nature of mechanical shocks, and since the optimal value of the DC voltage across the storage element depends on the shock amplitude, realizing this impedance matching is a hard task. The fact that SECE efficiency does not depend on the voltage level across the piezoelectric generator, and hence does not require impedance matching, seems promising in the case of shocks. Furthermore, new nodes technologies tend to lower the required voltage to power the sensors, which makes the SSHI interface even less efficient. For all those reasons, we chose to focus on the SECE interface.

III. SHOCK-OPTIMIZED SECE MODELLING

A. Reminder of SECE strategy

The SECE strategy has been widely investigated as one of the major piezoelectric energy harvesting strategy [11,18]. In previous art, it has been implemented various times using discrete components [18,23] or with a dedicated ASIC [24,25]. Here, we will remind the operation principle of the SECE strategy, and analyze its electromechanical efficiency when associated with shocks excitations. Fig.7 shows the displacement and piezoelectric voltage waveforms while harvesting the energy using SECE strategy.

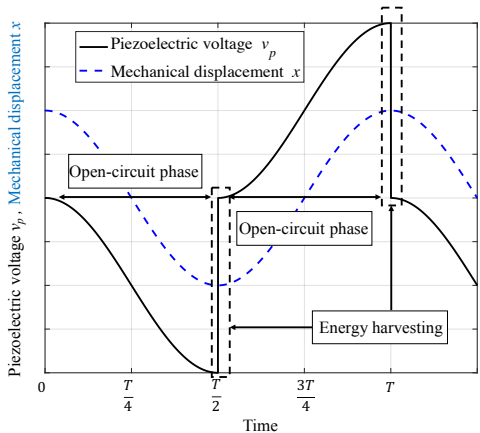

Fig.7. Waveforms associated with the SECE interface

The SECE strategy is a two-phase harvesting process: during most of the semi-period of vibration, the piezoelectric material is in open-circuit configuration. All the charges generated thanks to the direct piezoelectric effect are stored in its clamped piezoelectric capacitance $C_p$. When the voltage reaches an extremum (corresponding as well to a mechanical displacement extremum), the stored energy in the piezoelectric material is hence maximal. At this particular moment, the energy is quickly extracted from the material. This is usually done by connecting an inductor to the piezoelectric material. Indeed, this forms a LC tank which starts resonating at a frequency way more important than the vibration's one. When there is no more electrostatic energy in the piezoelectric material ($v_p = 0$), a new open-circuit phase starts. The electrical energy stored in the inductor is then transferred to the storage element. This energy, extracted from the piezoelectric material every semi-period of vibration, can be expressed by (6):

$$E_{elec} = \frac{1}{2} C_p V_m^2 \qquad (6)$$

With $V_m$ being the voltage across the piezoelectric material when the energy starts being extracted.

B. Analysis of the SECE under shocks

In order to predict the efficiency of a harvester coupled with a non-linear interface, numerical methodologies have already been developed, as extensively detailed in [26]. In this part, we will propose for the first time an analytical model which predicts the efficiency under shocks of the SECE strategy with any piezoelectric harvester. This model is based on the discretization of the previously given energy balance equation (2). In order to simplify the calculations, we will consider that all the energy contained in the shock happening at $t_0 = 0$ is stored in a very quick time in the stiffness of the mechanical resonator. This assumption is true as long as the mechanical losses between the shock occurrence and the first displacement extremum remain small. In that case, the energy contained in the shock can be expressed as:

$$E_{in} = \frac{1}{2} K_{sc} X_{m_0}^2 \qquad (7)$$

Where $X_{m_0}$ is the first displacement extremum of the tip mass after the shock. In order to solve (2), we assume that the quality factor of the resonator is much higher than 1. This is usually the case for commercial cantilever-based harvesters, whose $Q_m$ are usually in the $20 - 100$ range. This means that even though the piezoelectric voltage is nonlinear, thanks to the filtering effect of the resonator, only the first harmonic of the voltage has a non-negligible impact on the harvester dynamics. We will consider as well that the displacement amplitude remains constant on every semi-period of vibration as expressed by (8).

$$x(t) = X_{m_i} \cos(\omega_1 t), \forall t \in [\frac{i\,\pi}{\omega_1}, \frac{(i+1)\,\pi}{\omega_1}] \qquad (8)$$

Where $\omega_1$ is the angular open-circuit resonant frequency of the harvester, given by $\sqrt{(k_m^2 + 1)K_{sc}M^{-1}}$, $X_{m_i}$ is the amplitude of the displacement during the $i^{th}$ semi-period of vibration ($i \in \mathbb{N}$). On any semi period of vibration, as shown in Fig. 7, the harvester works in open circuit, thus there is no piezoelectric current $i_p$ flowing out of the piezoelectric material. From $i_p = 0$, (1) and (6), $E_{elec_i}$ can be expressed as (9).

$$E_{elec_i} = \frac{C_p}{2}\left[\int_{\frac{i\pi}{\omega_1}}^{\frac{(i+1)\pi}{\omega_1}} -\omega_1 \frac{\alpha X_{m_i}}{C_p}\sin(\omega_1 t)dt\right]^2 \quad (9)$$
$$= \frac{2\alpha^2 X_{m_i}^2}{C_p}$$

From (7) and (9), with $t = \frac{\pi}{\omega_0}$ and $i = 0$, and considering that the kinetic energy is null because all the mechanical energy is stored in the resonator's stiffness, we can evaluate the elastic potential energy in the system after a semi-period of vibration, given by $E_{pot_1}$:

$$E_{pot_1} = E_{in} - \frac{D\pi\omega_1}{2}X_{m_0}^2 - \frac{2\alpha^2}{C_p}X_{m_0}^2 \quad (10)$$

$\frac{D\pi\omega_1}{2}X_{m_0}^2$ modelling the energy losses due to mechanical damping $E_{damp}$ between the first and second displacement extrema. We can thus inject (7) in (10), and find $E_{pot_1}$:

$$E_{pot_1} = E_{in}\left(1 - \frac{D\pi\omega_1}{K_{sc}} - \frac{4\alpha^2}{K_{sc}C_p}\right) \quad (11)$$

Considering that the remaining potential energy in the mechanical system is $E_{pot_1}$, applying again the energy balance to the system yields (12).

$$E_{pot_1} = E_{pot_2} + \frac{D\pi\omega_1}{2}X_{m_1}^2 + \frac{2\alpha^2}{C_p}X_{m_1}^2 \quad (12)$$

We eventually find the recurrence expression of the potential energy $E_{pot_{i+1}}$ after $(i+1)$ semi period of vibration as a function of $E_{pot_i}$, mechanical potential energy after $i$ semi-period of vibration.

$$E_{pot_{i+1}} = E_{pot_i}\left(1 - \frac{D\pi\omega_1}{K_{sc}} - \frac{4\alpha^2}{K_{sc}C_p}\right) \quad (13)$$

Thus, (13) confirms that the mechanical energy in the system is decreasing, due to both the mechanical losses and the harvested energy. From (13), we can get the expression of $E_{pot_i}$ as a function of the initial energy in the system.

$$E_{pot_i} = E_{in}\left(1 - \frac{D\pi\omega_1}{K_{sc}} - \frac{4\alpha^2}{K_{sc}C_p}\right)^i \quad (14)$$

When the mechanical displacement reaches another extremum, all the mechanical energy is stored in the resonator's stiffness, as expressed by (15).

$$E_{pot_i} = \frac{1}{2}K_{sc}X_{m_i}^2 \quad (15)$$

From (14) and (15), we can find the mechanical displacement amplitude of the $i^{th}$ displacement extrema (16).

$$X_{m_i} = \sqrt{\frac{2E_{in}}{K_{sc}}\left(1 - \frac{D\pi\omega_1}{K_{sc}} - \frac{4\alpha^2}{K_{sc}C_p}\right)^i} \quad (16)$$

Combining (6), (9) and (16), the expression of the $i^{th}$ voltage extrema can also be determined, as shown by (17).

$$v_{m_i} = \frac{2\alpha}{C_p}\sqrt{\frac{2E_{in}}{K_{sc}}\left(1 - \frac{D\pi\omega_1}{K_{sc}} - \frac{4\alpha^2}{K_{sc}C_p}\right)^i} \quad (17)$$

From (9) and (16), the harvested energy during a single semi-period of vibration can be expressed by (18).

$$E_{elec_i} = \frac{4\alpha^2}{C_p K_{sc}}E_{pot_i} \quad (18)$$

Combining (14) and (18) and summing the result, we can get the total energy harvested by the SECE interface.

$$E_{elec} = \sum_{i=0}^{+\infty}\frac{4\alpha^2}{C_p K_{sc}}E_{in}\left(1 - \frac{D\pi\omega_1}{K_{sc}} - \frac{4\alpha^2}{K_{sc}C_p}\right)^i \quad (19)$$

Thus, applying geometrical series sum rules, we can get the expression of the harvested energy given by (20).

$$E_{elec} = \frac{4E_{in}\alpha^2}{C_p K_{sc}\left(\frac{D\pi\omega_1}{K_{sc}} + \frac{4\alpha^2}{C_p K_{sc}}\right)} \quad (20)$$

Considering the normalized electromechanical variables $k_m^2$ and $Q_m$, we can get an efficiency expression of the SECE interface.

$$\eta_{electromech} = \frac{E_{elec}}{E_{in}} = \frac{1}{\frac{\pi\sqrt{1+k_m^2}}{4k_m^2 Q_m} + 1} \quad (21)$$

The obtained expression is quite simple. $\eta_{electromech}$ does not depend on the voltage across the storage element nor on the shock amplitude, which is coherent with the supposed robustness of the SECE strategy. For our harvester (Table I), we have $Q_m = 92$ and $k_m^2 = 0.8\%$, leading to a theoretical electromechanical efficiency of 48.2%, which is in agreement with the numerical simulation (48%) shown in Fig.5.

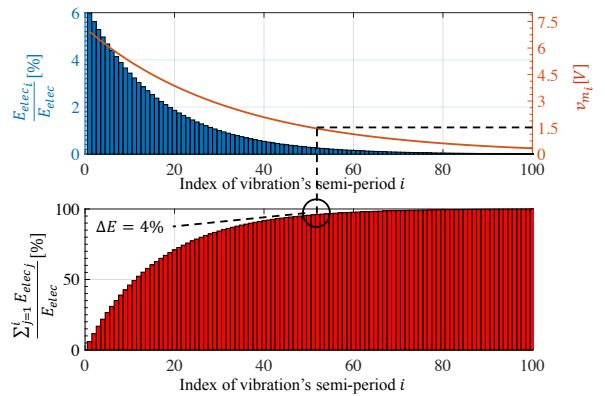

Fig.8. Electrical energy, piezoelectric voltage amplitude, and cumulative harvested energy during a single shock with $E_{in} = 22\mu J$

Thanks to (17) and (18), we have been able to compute the voltage and harvested energy at each displacement extremum, as shown in Fig.8. When the piezoelectric voltage is below 1.5V, we can observe from Fig.8 that the difference between the harvested energy and total energy is relatively small, around 4% of the total energy. Thus, we choose not to use the SECE strategy when the piezoelectric voltage is below 1.5V in order not to waste energy in powering the different control blocks. That is the reason why the proposed system only starts if the piezoelectric voltage reaches 1.5V, as explained extensively in the next part.

## IV. ARCHITECTURE OF THE PROPOSED SHOCK-OPTIMIZED SECE

The proposed harvesting circuit can be observed in Fig.9. The circuit is composed of a negative voltage converter (NVC) rectifying the PEH output voltage, and a SECE power path controlled by a sequenced circuit.

### A. Cold start

The voltage supplying the whole interface circuit, $V_{ASIC}$, is stored in an external capacitance $C_{ASIC}$, which is connected to $C_{STORE}$ thanks to a PMOS diode. If $V_{ASIC}$ is below $1.5V$, the energy stored in $C_{ASIC}$ is considered too low to ensure the control circuit self-operation. The harvested energy is then transferred from the piezoelectric harvester to $C_{ASIC}$ thanks to a non-optimized path. This path is represented in Fig.9 as a NVC followed by the cold start block, and works exactly as a SEH interface. As soon as the voltage across $C_{ASIC}$ reaches $1.5V$, the sequencing starts. This sequencing is divided in 4 phases $T_1 - T_4$, and is described extensively in the following.

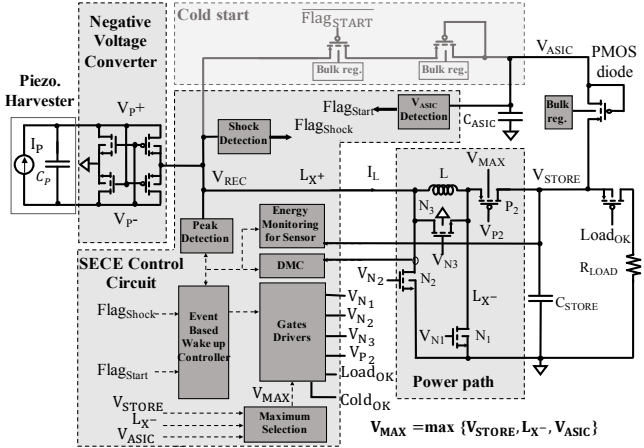

Fig.9. Proposed Shock-optimized SECE circuit

### B. Sleeping mode, $T_1$

A system-level view of our system can be observed on Fig.10, where only the harvester, the NVC and the power path are shown. At first, when there is no shock, or when the absolute voltage across the piezoelectric element is below $V_{TH}(1.5V)$, the circuit is in sleeping mode, $T_1$. In order to minimize the static consumption, all blocks except the Shock Detection (SD) block are turned off.

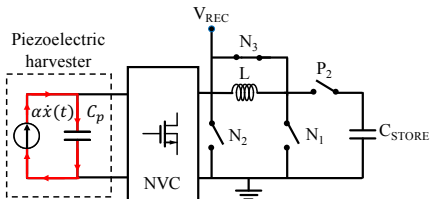

Fig.10. Proposed harvesting circuit during $T_1$

When a shock happens, the voltage across the piezoelectric material starts increasing, as well as the rectified piezoelectric voltage $V_{REC}$. Next, the SD checks if the electrical energy converted by the piezoelectric transducer is sufficiently high to be harvested ($V_{REC} > V_{ASIC}$). By setting Flag$_{Shock}$, the SD enables the $V_{ASIC}$ Detection which determines whether or not the cold start path should be activated. If $V_{ASIC}$ is below $V_{TH}$, we consider that the stored energy is insufficient to start the SECE operation, and the cold start path remains connected in order to keep charging $C_{ASIC}$. If this is not the case, the $V_{ASIC}$ Detection sends the Flag$_{Start}$ signal which disables the cold start, turns on the Peak Detection block, and starts the maximum voltage detection phase $T_2$, depicted in Fig.11.

### C. Maximum voltage detection phase, $T_2$

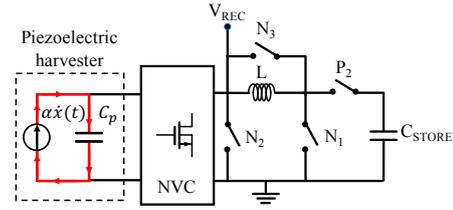

Fig.11. Proposed harvesting circuit during $T_2$

As soon as $V_{REC}$ reaches an extremum, as shown in Fig.7, the system enters its harvesting phase, $T_3$, as shown in Fig.12. The Peak Detection is then turned off, while the Dual Mode Comparator (DMC) in its Zero Crossing Detection (ZCD) is enabled. $V_{N1}$ is set high in order to close $N_1$, which connects the inductance L with the piezoelectric capacitance $C_p$.

### D. Harvesting phase, $T_3$

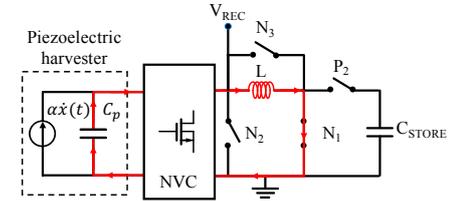

Fig.12. Proposed harvesting circuit during $T_3$

As the energy is transferred from the piezoelectric capacitance $C_p$ to L, the voltage across the piezoelectric quickly decreases. When $V_{REC}$ goes below a voltage boundary called $V_{TL}$=-14mV, meaning that all the energy has been transferred from $C_p$ to L, the ZCD sends a signal, and is turned off. The DMC is shifted to its Reverse Current Detection (RCD), and $N_1$ is open while $N_2$ and $P_2$ are simultaneously closed. Thus, the inductance L is now connected to the storage capacitance $C_{STORE}$, as the circuit starts its storing and final phase $T_4$.

### E. Storing phase, $T_4$

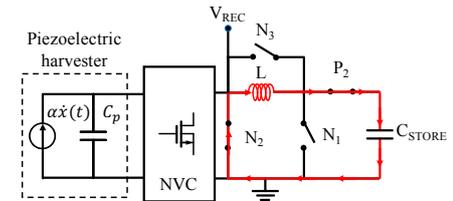

Fig.13. Proposed harvesting circuit during $T_4$

The current in the inductance L starts decreasing as the energy is transferred from L to $C_{STORE}$. When $I_L$ reaches $0A$, the RCD sends a flag to the system, indicating that the energy has been transferred successfully from the piezoelectric capacitance to the storage one. Ultimately, the circuit returns to its sleep mode $T_1$, waiting for the next harvestable energy event. $N_3$ acts as a freewheeling diode, and provides a path to dissipate the unavoidable remaining energy in L.

*F. Sequencing summary*

The control circuit state machine and sequencing are shown in Fig.14 while the waveforms and chronograms associated with the proposed sequencing are shown in Fig.15.

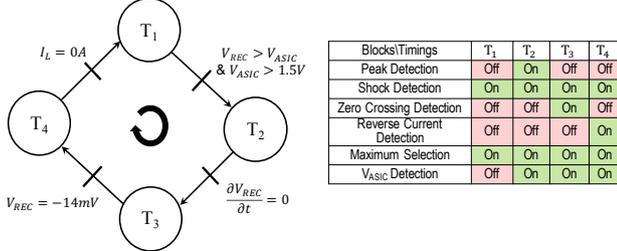

Fig.14. State machine and sequencing of the proposed control circuit

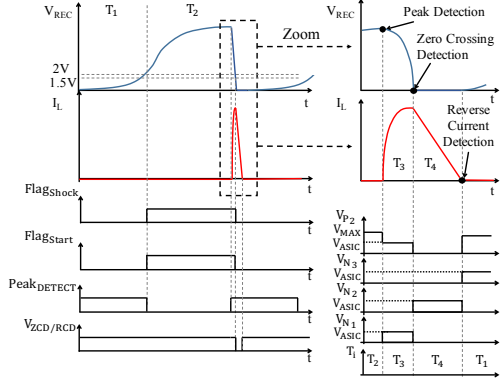

Fig.15. Waveforms and chronograms of the proposed control circuit

## V. CIRCUIT IMPLEMENTATION

In this part we show and describe the building blocks of the shock-optimized SECE system. In the following, the threshold voltage of the transistors, noted $V_T$, is equal to 0.45V.

*A. Shock and $V_{ASIC}$ Detections*

As explained extensively in the previous part, in order to transition from $T_1$ to $T_2$, both the voltage across the PEH and the voltage across $C_{ASIC}$ should be important enough. In this perspective, a Shock Detection and a $V_{ASIC}$ Detection blocks are sequentially used. Their transistor-level implementations are shown in Fig.16. The resistances $R_0$, $R_1$, $R_2$, $R_3+R_{33}$, and $R_4$, have been fixed to 2MΩ, 2MΩ, 8MΩ, 22.4MΩ, and 9.1MΩ, respectively.

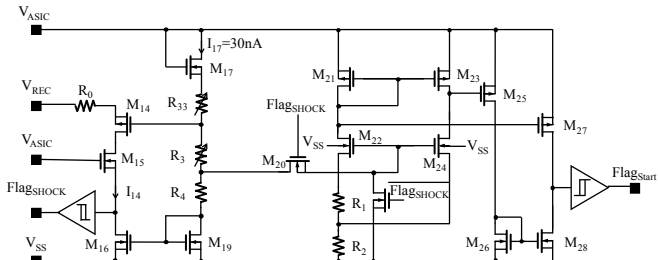

Fig.16. Shock Detection and $V_{ASIC}$ Detection schematics

During $T_1$, the only current drawn from $C_{ASIC}$, named $I_{17}$, whose expression is given by (22).

$$I_{17} = \frac{\left(V_{ASIC} - V_{gs_{17}} - V_{gs_{19}}\right)}{R_{33} + R_3 + R_4} \quad (22)$$

When $V_{REC}$ grows, the current $I_{14}$ flowing through $M_{14}$ is consequently increased. When $I_{14}$ becomes equal to $I_{17}$, the current mirror formed by $M_{16}$ and $M_{19}$ is balanced, and $M_{16}$ drain potential is high. This consequently sends a Flag$_{SHOCK}$ signal to the state machine (Fig.14). The condition for sending Flag$_{SHOCK}$ can be written as:

$$V_{REC} > V_{ASIC} + V_{gs_{14}} - V_{gs_{17}} - I_{17}(R_{33} - R_0) \quad (23)$$

The threshold value of $V_{REC}$ respecting condition (23), $(V_{REC})_{th}$, is almost insensitive to temperature fluctuations thanks to the mutual cancellations of $M_{14}$ and $M_{17}$ gate-source voltages and the negligible value of $I_{17}$ term compared to $V_{ASIC}$. The programmable resistive divider ($R_3$ and $R_{33}$) allows to adjust $(V_{REC})_{th}$ value from which we consider worth to enable the active path. In our case, when we choose to have $R_{33}=0Ω$ and $V_{ASIC}=2V$, we obtain $I_{17}=35nA$ and $(V_{REC})_{th}≈2V$.

When Flag$_{Shock}$ becomes high, this consequently enables the $V_{ASIC}$ Detector by forcing $M_{20}$ conduction. The integrated resistances $R_1$ and $R_2$ enable the minimum $V_{ASIC}$ to be selected, to ensure the self-operation of the chip. In our case, we fixed this minimum $V_{ASIC}$ at 1.5V. When this condition is satisfied, Flag$_{Start}$ is set high thanks to a two stage comparator, which allows the transition from $T_1$ to $T_2$.

*B. Peak Detection*

As soon as the system enters in $T_2$, the PD is enabled in order to detect $V_{REC}$ extremum. The schematic of this peak detection block is illustrated in Fig.17. Since $V_{REC}$ is increasing, the current flowing through $C_0$, $i_{PD}$, is positive. $M_{31}$ is blocked while $M_{32}$ acts as a current sink. The Common Source Stage (CSS) made of two transistors, $M_{37}$ and $M_{36}$ amplifies $V_{PKIN}$ to drive $M_{31}$ and $M_{32}$ gates. As $V_{REC}$ gets closer to its extremum (right before $T_3$), the current $i_{PD}$ gets closer to 0A. $V_G$ gets larger which progressively blocks $M_{32}$ and starts $M_{31}$ thanks to CSS feedback, while $V_{PKIN}$ tends to decrease. At the instant the current polarity changes, $V_G$ is set high. However, $V_{PKIN}$ is not maintained constant, which implies a lagging between $V_{REC}$ maximum detection and $i_{PD}$ zero crossing, as expressed by (24).

$$i_{PD} = 0 \rightarrow \frac{\partial(V_{REC})}{\partial t} = \frac{\partial(V_{PKIN})}{\partial t} < 0 \quad (24)$$

Thus, there is an intrinsic delay between $V_G$ transition and $V_{REC}$ extremum. This delay, named $t_{dv}$, is inversely proportional to the gain of the CSS. The step on $V_G$ is amplified thanks to another CSS made of $M_{40}$ and $M_{41}$. The intrinsic angular lagging between the output signal of PD and $V_{REC}$ extremum $\Delta t$ is the sum of $t_{dv}$, $t_{leak}$, and $t_{dyn}$, which are respectively the delay induced by $\frac{\partial(V_{PKIN})}{\partial t} \neq 0$, the delay induced by the leakage currents in $M_{31}$ and $M_{32}$, and the delay induced by the two CSS.

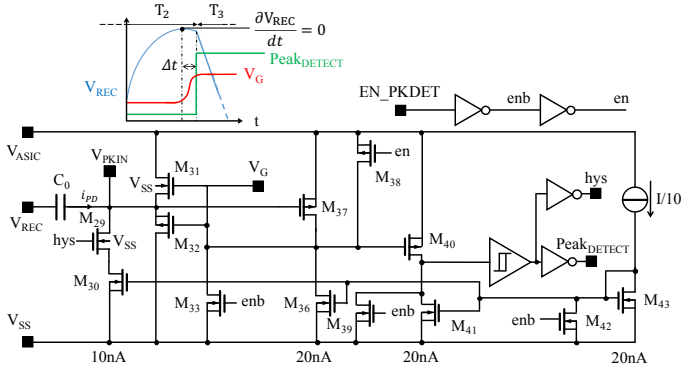

Fig.17. Peak detection schematic

In transistor-level simulations with both process variations (Montecarlo), temperatures going from -40°C to +120°C, and frequencies ranging from 30Hz to 200Hz, the maximum angular delay $\Delta\phi = \omega_1 \Delta t$ is 5.2 degrees. With this lagging, less than 1% of the energy is lost compared when $\Delta\phi = 0$. This proves that the lagging induced by the peak detection block has, for our PEH, negligible impact on the extracted power.

### C. Dual Mode Comparator

In $T_3$, the ZCD is enabled to detect the moment the voltage goes below $0V$. Figure 18 shows the transistor-level implementation of a Dual Mode Comparator (DMC) which is used in $T_3$ as a ZCD. $M_2$ and $M_4$ constitute a differential pair allowing $V_{REC}$ to be compared with the ground voltage. Due to $M_1$, when $V_{REC}$ is high, only ¼ of the bias current flows through $M_1$ and $M_2$. As $V_{REC}$ decreases (thanks to the charge transfer occurring between $C_P$ and L), the current in $M_2$ is increased, which improves the detection accuracy. Furthermore, the circuit consumption is reduced when $V_{REC}$ is high, since it is only useful to increase the comparator performances when $V_{REC}$ gets close to 0V. $V_{HYST}$ is initially high, which creates an offset on the input of the comparator since the two resistances connected to $M_2$ and $M_4$ are not identical. The value of this offset should be below 0V in order to generate a hysteresis, but it should not be too low, in order not to waste too much energy. In our case, we fixed this offset to $-14mV$, by choosing the unit resistance $R = 62k\Omega$ and the unit current $I = 200nA$. When $V_{REC}$ reaches $-14mV$, $V_{HYST}$ is set to $0V$ and the DMC sends a flag to the system indicating the transition from $T_3$ to $T_4$.

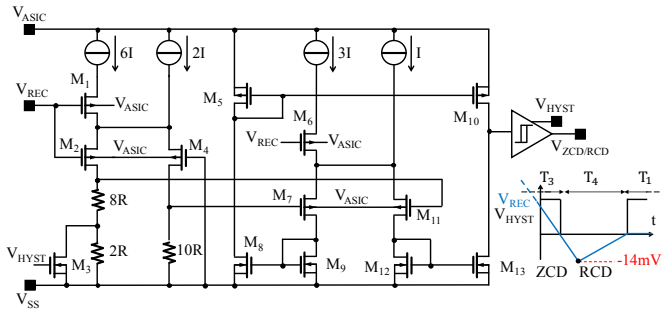

Fig.18. Dual Mode Comparator schematic

In $T_4$, the DMC switches to its RCD configuration. As $V_{HYST}$ set low, there is no more a -14mV offset on the comparator input. In this phase, $V_{REC}$ is proportional to $-I_L$, as $N_2$ is turned on. Therefore, when $I_L$ decreases, $V_{REC}$ increases until it reaches 0V. Then, $T_1$ starts again: the DCM is disabled in order to avoid any unnecessary energy consumption, and only the SD is powered.

### D. Bias Generator

The bias generator schematic shown in Fig.19 is used between $T_2$ and $T_4$ in order to provide regulated biasing currents from 20nA to 200nA. $M_{53}$, $M_{55}$, $M_{51}$ and $M_{58}$ form a current PTAT generator. The current flowing in those transistors is fixed to 40nA thanks to $R_{bias}$. In order to start the bias, a starter made of $M_{48}$, $M_{47}$, and a MOS resistance ($M_{44}$-$M_{46}$) has been designed. When the enable signal $EN_{bias}$ is set high (in $T_2$), $V_{START}$ is initially equal to $V_{ASIC}$. The current starting to flow through $M_{48}$ is around 3μA, and will quickly lower the gate voltages of $M_{51}$ and $M_{58}$. This will force the conduction of $M_{51}$, $M_{58}$, and ensure the quick starting bias of $M_{53}$ and $M_{55}$. The current flowing through $M_{53}$ is recopied in $M_{47}$, which draws $V_{START}$ to the ground, disabling the start-up circuitry. In most case, the time to enter in steady-state is around a few μs, which is much shorter than $T_2$.

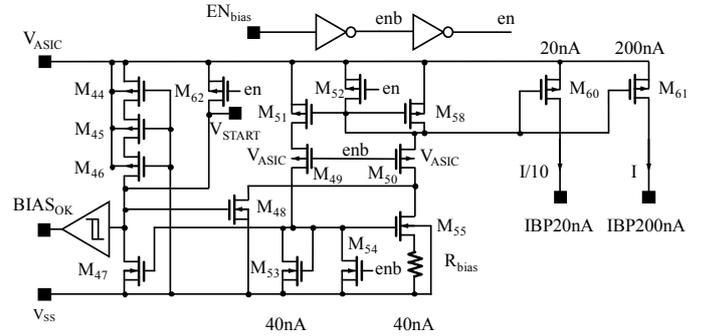

Fig.19. Bias generator schematic

### E. $P_2$ gate driver and maximum selection

A particular attention has been given to the gate driver of transistor $P_2$, illustrated in Fig.20. $P2_{ON}$ is a digital signal generated by the event based wake up controller. When $P2_{ON}$ is high, the gate voltage of $P_2$, $V_{P2}$, is low (during T4). When $P2_{ON}$ is low, $V_{P2}$ is high (during T1, T2, T3), and its value depends on $P2GATE_{LVL}$, a digital signal coming from the event based wake up controller that fixes the inverter chain's supply voltage. When $P2GATE_{LVL}$ is high, the gate voltage of $P_2$ is set to $V_{MAX}$ (during T1 and T2) in order to block any current through $P_2$. In $T_3$, $V_{P2}$ is set to $V_{ASIC}$ instead of $V_{MAX}$. Indeed, in the transition from $T_3$ to $T_4$, the transistor $N_1$ is turned off, but we want to keep on providing a path for the inductance current, and avoid an important increase of $L_{X^-}$. Thus, since $V_{P2}$ is set to $V_{ASIC}$, $L_{X^-}$ is clamped to $V_{ASIC}+V_T$, and $P_2$ is forced to conduct the current from the inductance to the storage capacitance even when there is a delay in $P2_{ON}$ signal. The output of the maximum selection block $V_{MAX}$ is given by $V_{MAX}$ = max ($V_{STORE}$, $L_{X^-}$, $V_{ASIC}$) and is generated thanks to $M_{73}$-$M_{76}$.

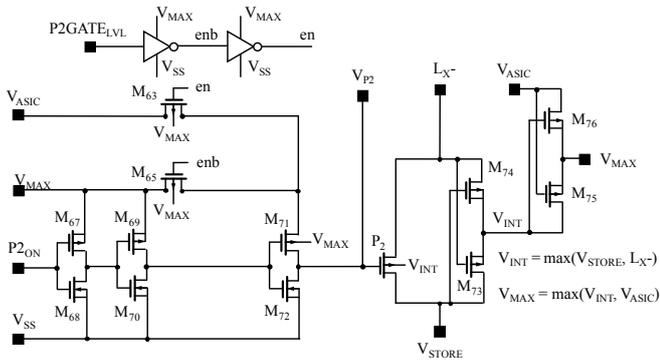

Fig.20. $P_2$ gate driver and maximum selection block schematics

*F. Energy Monitoring for Sensor*

The Energy Monitoring block can be activated manually with a digital pin that can be connected to $V_{ASIC}$. The Energy Monitoring block is made of a comparator which sets high the signal $Load_{OK}$ as soon as the voltage in the storage capacitance $V_{STORE}$ is greater than 2.75V. When $Load_{OK}$ is high, the energy is transferred from $C_{STORE}$ to the sensor node, and consequently, $V_{STORE}$ decreases. As soon as $V_{STORE}$ gets below 2.25V, $Load_{OK}$ is set to 0V, and the energy stops being transmitted to the sensor. The demonstration of this Energy Monitoring block is shown in the experimental part, in Fig.22.

*G. PCB and chip*

Our chip was fabricated in CMOS 40nm triple-well technology including 10V devices, and occupies a 0.55mm² core area. Our PCB and chip micrograph are shown in Fig.21.

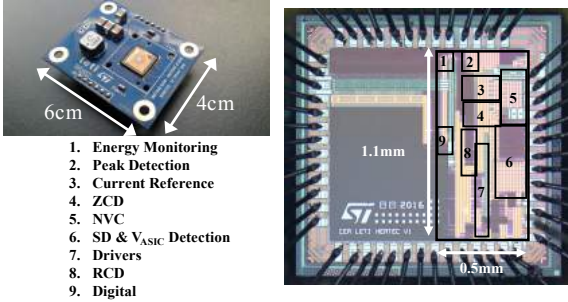

1. Energy Monitoring
2. Peak Detection
3. Current Reference
4. ZCD
5. NVC
6. SD & $V_{ASIC}$ Detection
7. Drivers
8. RCD
9. Digital

Fig.21. PCB and Chip micrograph

## VI. EXPERIMENTAL RESULTS

*A. Experimental setup*

Experimentations have been conducted. In order to emulate both periodic and shock excitations, a MIDE piezoelectric generator (PPA1011) with a 5.67g mobile mass and a resonant frequency of 75.4 Hz has been placed on a shaker. The characteristics of the piezoelectric harvester have already been summarized in Table I. A PCB circuit has been designed in order to interface our ASIC and includes an off-chip inductance L of $0.13 cm^3$ and two capacitances $C_{STORE}$ and $C_{ASIC}$, whose values are 2.2mH, 100μF, and 10μF, respectively.

*B. Experimental waveforms*

Fig.22 shows both the voltage across the piezoelectric harvester and the storage capacitance when shocks are applied every second on the harvester. After the first three shocks, which are used to store enough energy in $C_{ASIC}$ thanks to the cold start power path, the system operates autonomously in its optimized mode and the energy is stored in $C_{STORE}$. For test purposes, when $V_{STORE}$ reaches 2.8V, the energy monitoring block intermittently connects a 1kΩ load resistance to emulate the power consumption of a sensor. In this setup and in steady-state, 10 shocks are required to reach enough energy to power the emulated sensor.

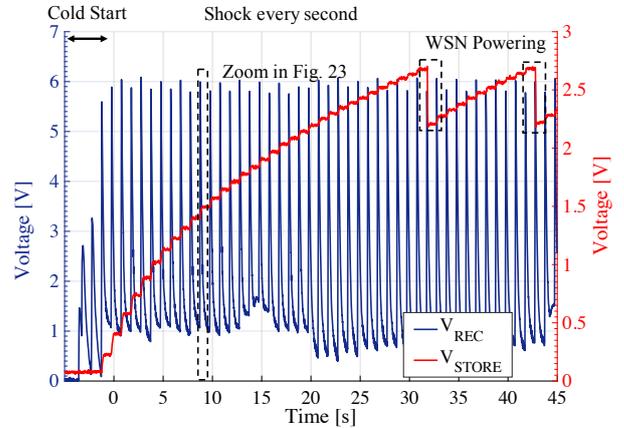

Fig.22. Experimental waveforms of the piezoelectric harvester under shocks

If we take a closer look at these waveforms for one single shock, as illustrated in Fig.23, we observe that the voltage amplitude is decreasing as the mechanical energy is both harvested and dissipated due to friction, which is consistent with the theoretical part.

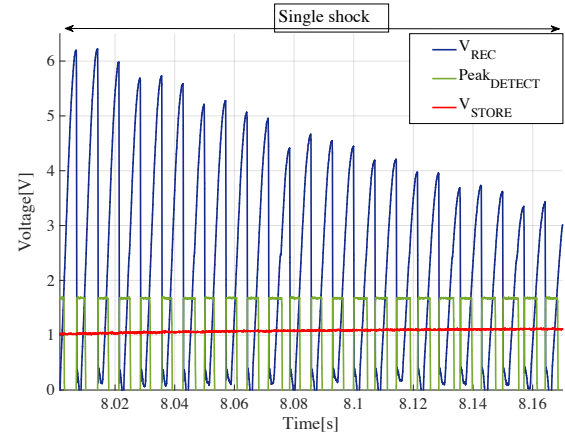

Fig.23. Experimental waveforms of the piezoelectric harvester under a single shock (zoom of Fig.22)

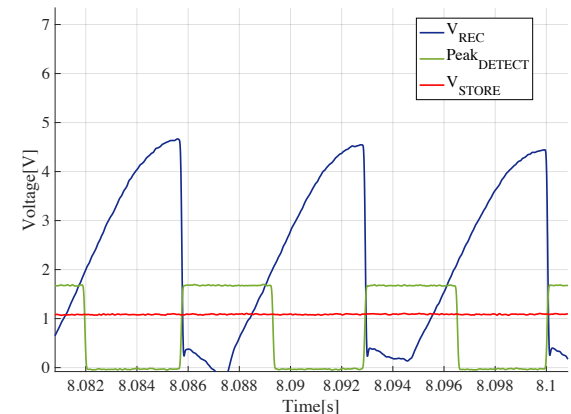

Fig.24. Experimental waveforms of the piezoelectric harvester for a few oscillations (zoom of Fig.23)

If we zoom once again on a few oscillations, as shown in Fig.24, we observe the SECE voltage waveforms, which correspond to the theoretical ones in Fig.7. The Peak$_{DETECT}$ flag sent by the PD is sent accurately when V$_{REC}$ reaches its extremum value, as shown in Fig.24.

*C. Results and Discussions*

The harvested power has been measured for various shock amplitudes and V$_{STORE}$. In order to experimentally determine the energy of each mechanical shocks $E_{in}$, we measured the first mechanical displacement extremum of the cantilever beam $X_{m_0}$, and calculated the energy contained in the shock using (7). To measure $E_{elec}$, we applied $E_{elec} = \sum_{i=1}^{\infty} \frac{1}{2} C_p V_{m_i}^2$, with $V_{m_i}$ being the measured piezoelectric voltage right before the $i^{th}$ energy harvesting event. In order to change V$_{STORE}$, we connected various values of resistance to the storage capacitance C$_{STORE}$ then measured the voltage across the latter. We compared it to the harvested power using the SEH interface. The SEH interface has been implemented by the NVC followed by a diode (1N4148) in order to block reverse currents [24], and connected to the storage capacitance.

The measured harvested power under $70\mu J$ shocks occurring every second are shown in Fig.25. We observe that the proposed shock-optimized SECE interface harvests 4.2x more maximum energy than the SEH interface. The harvested power of the SECE remains important and almost constant as long as the voltage across the storage capacitance V$_{STORE}$ remains between 1V to 3V. This confirms the robustness of the SECE over the voltage range that should be used in order to power various sensor nodes. The extracted electrical energy $E_{elec}$ has been measured and is in good agreement with the theoretical predictions given by (20), with less than 8% error. The differences between experimental results and analytical modelling are mainly due to the threshold voltage $V_{TH}$ required to start the self-operating SECE, as highlighted in Fig.8. The experimental electromechanical efficiency $\eta_{electromech}$, electrical efficiency $\eta_{elec}$, and end-to-end efficiency $\eta_{end-to-end} = \eta_{electromech} \cdot \eta_{elec}$ under the same $70\mu J$ shocks are shown in Fig. 26. The maximum electrical efficiency $\eta_{elec}$ of our circuit under shock reaches 91% for a V$_{STORE}$ equals to $2.38V$.

Fig.27 shows the harvested power under smaller shocks of $20\mu J$. The maximum harvested power of our interface is still at least three times superior of the SEH's one. We can observe that the gain of the SECE interface is a bit smaller. Indeed, as the voltage across the piezoelectric harvester gets smaller, the efficiency of our interface consequently decreases.

The FoM of our circuit, previously defined in [14, 25], is expressed by (25).

$$FoM = \frac{\max(P_{SECE})}{\max(P_{SEH})} \quad (25)$$

Our circuit FoM has been measured for various shocks energies, and is given in Fig.28. As the shocks happen every second, the harvested energy during a single shock $E_{stored}$ can be directly converted in power, using $P_{SECE} = E_{stored}$. We can observe that for all energy-range shocks, our circuit remains at least 2.5 times more efficient than the SEH interface. Under $70\mu J$ shocks, our circuit is 4.2 times more efficient than the SEH, which corresponds to the maximum of our interface's $FoM_{SHOCK}$. When the shock energy is increased, the $FoM$ starts decreasing. Indeed, while the SECE interface efficiency remains relatively constant, the SEH interface becomes more and more efficient as the shock energy is increased, because the diode threshold voltage and MOS $V_T$ become negligible compared to the voltage across the piezoelectric element. Furthermore, the fact it does not require any impedance matching circuit which would include additional losses make it even better.

Thanks to the sequencing of our circuit, the measured quiescent current in sleeping mode is 30nA@1.5V. This allows self-operation of our circuit with an input electrical power as low as 80nW, which is considerably lower than what can be found in previous art. This low quiescent current allows our circuit to maintain self-operation under very harsh conditions, i.e. under small shocks producing $8\mu J$ of electrical energy $E_{elec}$ happening every 100 seconds, as shown in Fig.29.

We measured the efficiency of our circuit as a function of the available electrical power, as shown in Fig. 30. To realize these measurements, the shock energy $E_{in}$ has been fixed to $70\mu J$, and V$_{STORE}$ was around 2V. In order to obtain various electrical power, we adjusted the time between every shock from 1 second to 375 seconds. When the time between every shock is approximately 370 seconds, the available electrical power is around 80nW, and the efficiency of our circuit tends to 0%. When the time between every shock gets close to 1s, the measured efficiency is around 90%, which is consistent with the results shown in Fig. 25 and Fig. 26 (when V$_{STORE}$ = 2V).

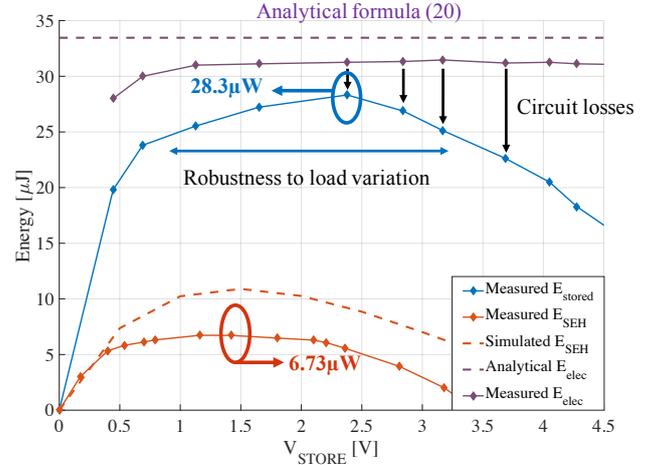

Fig.25. Experimental and theoretical measurement of the harvested power using SECE and SEH under $70\mu J$ shocks

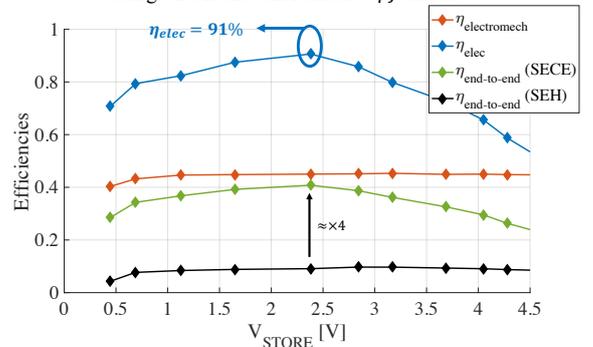

Fig.26. Experimental measurement of the electromechanical, electrical and end-to-end efficiencies under $70\mu J$ shocks

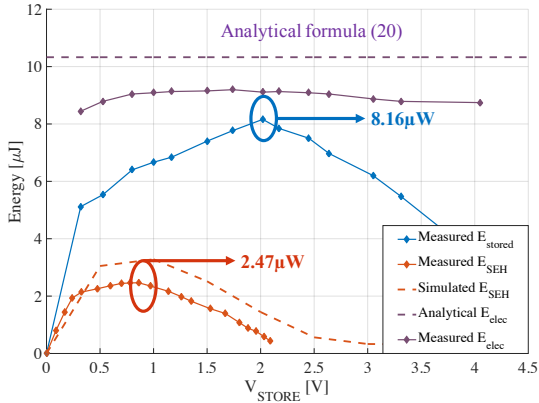

Fig.27. Experimental measurement of the harvested power using SECE and SEH under 20µJ shocks

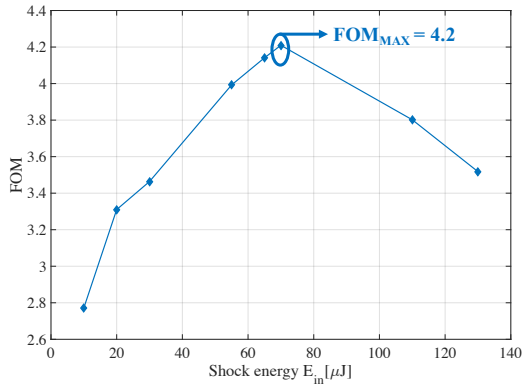

Fig.28. FoM of our circuit under various shock amplitudes

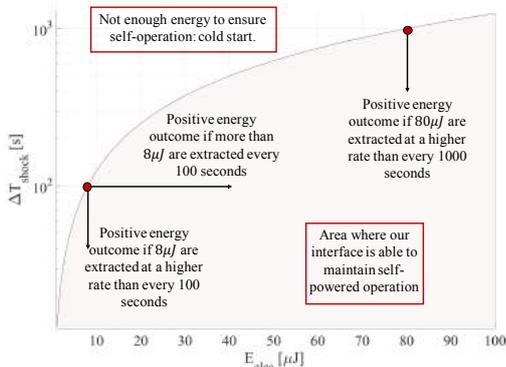

Fig.29. Theoretical self-operation limit of our circuit as a function of the extracted energies $E_{elec}$ and time between every shock $\Delta T_{shock}$.

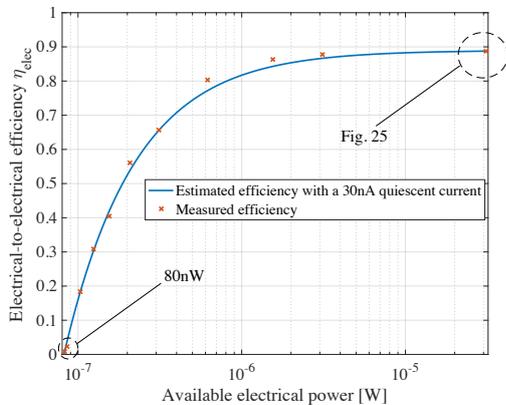

Fig.30. Measured electrical-to-electrical efficiency of the proposed SECE interface as a function of the electrical power

We also measured the performances of our circuit under periodic excitations of 1g at 75.4Hz, as shown in Fig.31.

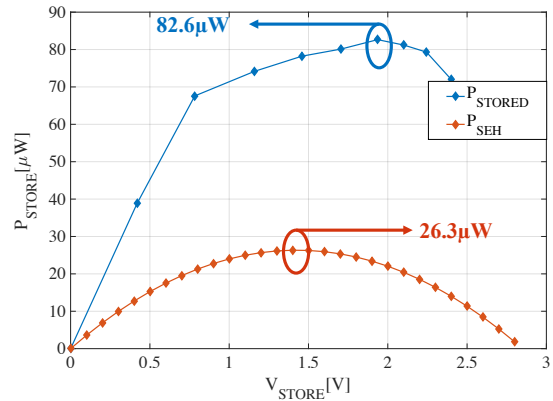

Fig.31. Experimental measurement of the harvested power using SECE and SEH under periodic excitation

Our circuit, even though it was not dedicated to harvest energy from periodic excitations, is still about three times more efficient than the SEH. Table II shows the comparison between our circuit and prior art. In comparison to previous work, our circuit exhibits a 1.6x FoM enhancement. The measured maximum electrical efficiency $\eta_{elec}$ of our circuit is 94% under periodic excitation at 82µW which is the highest end-to-end efficiency compared to former works. Compared to other SECE interfaces, our circuit performances under periodic excitations are also the best. However, for periodic excitations, SSHI interface remains the most interesting interface, especially for lowly coupled and highly damped piezoelectric harvesters.

TABLE II
PERFORMANCES COMPARISON WITH PRIOR ART

|  | [12] | [24] | [13] | [14] | This Work | Unit |
|---|---|---|---|---|---|---|
| Technology | 350 | 350 | 350 | 350 | 40 | nm |
| Chip Size | 3.6 | 1.25 | 2.34 | 0.72 | **0.55** | mm² |
| Scheme Type | SECE | SECE | Energy Investing | SSHI | SECE | - |
| Piezoelectric Harvester | Murata | MIDE V22B | MIDE V22B | MIDE V21B & V22B | MIDE PPA1011 | - |
| $C_P$ | 23 | 19.5 | 15 | 26 | 43 | nF |
| Excitation type | Periodic | Periodic | **Periodic & Shock** | **Periodic & Shock** | **Periodic & Shock** | - |
| Operation Frequency | 100 | 174 | 143 | 225 | 75.4 | Hz |
| FOM (periodic) | ≈170* | 206 | 360 | **681** | 314 | % |
| FOM (shocks) | N/A | N/A | - | 269 | **420** | % |
| Cold Startup | **Yes** | **Yes** | No | **Yes** | **Yes** | - |
| Elec.-to-elec. Efficiency $\eta_{elec}$ | 61 | 85.3 | 69.2 | Not available | **94 (periodic) 91 (shocks)** | % |
| Quiescent current | 1 | 0.3 | 0.1 | ≈1* | **0.03** | µA |

*Estimated from the data given in the paper

## VII. CONCLUSION

In this paper, we first explain the electromechanical model of a piezoelectric harvester under shock. We showed that the aim of the electrical interface, under shock excitations, is to harvest the energy from the mechanical resonator as quick as possible. We proposed a numerical comparison between different approaches to harvest energy from shock excitation, based on an energy balance modelling of the harvester. Coupled with our piezoelectric harvester, we showed that the SECE interface was the most appropriate one. Thereafter, thanks to the energy balance analysis, we derived an analytical expression of the extracted electrical energy using SECE strategy. The proposed interface has then been detailed, both at system and transistor level. The proposed IC in 40nm technology allows to add harvesting functionalities within a microcontroller die. The dedicated sequencing allows to have a low quiescent current of around 30nA and to maximize the electrical efficiency, up to 94% under periodic excitation. Compared to a SEH interface, the proposed interface harvests up to 420% more energy from shock excitations, which is, to the authors knowledges, the best shock FoM among state-of-the-art energy harvesting interfaces.

## Acknowledgments

This research was, in part, funded by the French Inter-ministerial Fund (FUI), through HEATec project, and by STMicroelectronics.

## Biographies

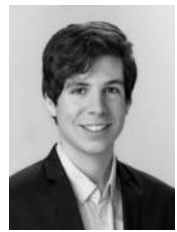

**Adrien Morel** was born in Valenciennes, France in 1993. He received his electrical engineering degree from the National Institute of Applied Sciences of Lyon (INSA Lyon) and his M.Sc. degree in integrated systems from Lyon University, both in 2016. He is currently pursuing his Ph.D at CEA Leti in Grenoble, France. His research interests are focused on micro energy harvesting, multiphysics interactions, integrated power management, and ultra-low-power analog circuit design.

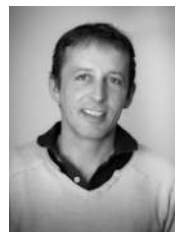

**Anthony Quelen** was born in Saint Adresse, France, in 1975. He received his Master degree in Electrical Engineering from University of Blois, France in 1999. From 2000 to 2005, he was involved in IC design development for audio applications in ON Semiconductor, Grenoble, France. From 2005 to 2014, he was technically responsible for ASIC design dedicated to power management for mobile phones in Maxim Integrated, Grenoble, France. In 2015, he joined CEA-Léti where he is in charge of analog IC design dedicated to integrated power supply and energy harvesting conditioning. His research focuses on sub-mW DC-DC converters and nW silicon-based voltage reference.

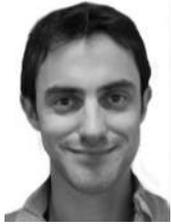

**Pierre Gasnier** received its Dipl. Ing. degree in electrical and electronic engineering from Polytech' Orleans, France in 2009; and its PhD degree from the University of Grenoble, France in 2014. From 2010 till 2013, he was working towards his PhD degree in the field of IC design and energy harvesting for Wireless Body Area Networks. Since 2013, he is with the CEA-LETI (System Division) as a research engineer. His research interests are mechanical energy harvesters and particularly vibration and flows, dedicated power management circuits and low-power electronics for Wireless Sensors Nodes.

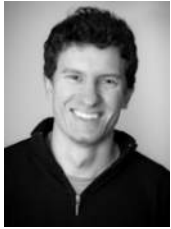

**Romain Grézaud** was born in Thonon, France, in 1987. He received the M.S. degree from the National Institute of Applied Sciences of Lyon INSA, France, in 2011. In 2014, he received the Ph.D. degree in electrical engineering from the Grenoble Institute of Technology INPG, France. He is currently a research scientist at the CEA-LETI in the Micro and Nanotechnology Innovation Centre MINATEC, France. His research interests include energy harvesting, power conversion and monolithic integration.

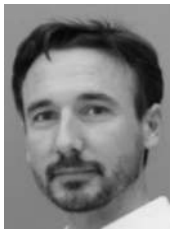

**Stéphane Monfray** received the M.Eng. degree in physics and the Postgraduate Diploma degree in microelectronics in 1999 from the Institut national des sciences appliquées de Lyon (INSA Lyon), Villeurbanne, France, and the Ph.D. degree from the Université de Provence Aix-Marseille I, Marseille, France, in 2003. In 1999, he joined France Telecom R&D, where he was engaged in the development and characterization of the silicon-on-nothing technology in collaboration with ST Microelectronics, Grenoble, France. He is currently a Disruptive Devices Project Manager in the Advanced Devices Group at ST Microelectronics. He was involved in the research of advanced devices integration for nine years. He was also involved in several European projects (Nanocmos, Pullnano), and driving and managing collaborations with universities, research laboratories, and supervises Ph.D. dissertations. He is the author and coauthor of more than 32 publications in major conferences and journals, of more than 20 patents, of a book chapter. Dr. Monfray was the corecipient of the Paul Rappaport Award in 2000. He had multiple participations and paper presentations at the International Electron Device Meetings in 2001, 2002, 2004, and 2007.

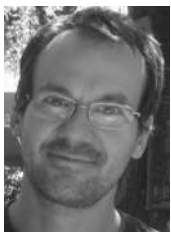

**Adrien Badel** graduated from Institut National des Sciences Appliquées de Lyon (INSA), Lyon, France, in electrical engineering in 2002 (MS degree). He prepared his Ph.D. at the Electrical Engineering and Ferroelectricity Laboratory of INSA Lyon, France. He received his Ph.D. degree in 2005 for his work on vibration control and energy harvesting. From November 2005 to November 2007, he was a JSPS (Japanese Society for the Promotion of Science) postdoctoral fellow at the Institute of Fluid Science of Tohoku University, Sendai, Japan. He is now a full professor at the Laboratory of Systems and Materials for Mechatronics from the Université de Savoie, Annecy, France. His research interests include energy harvesting, vibration damping and piezoelectric actuators modeling and control.

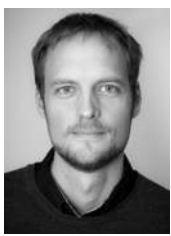

**Gaël Pillonnet** was born in Lyon, France, in 1981. He received his Master's degree in Electrical Engineering from CPE Lyon, France, in 2004, a PhD and habilitation degrees from INSA Lyon, France in 2007 and 2016, respectively. Following an early experience as analog designer in STMicroelectronics in 2008, he joined the University of Lyon in the Electrical Engineering department. During the 2011-12 academic year, he held a visiting researcher position at the University of California at Berkeley. Since 2013, he has been a full-time researcher at the CEA-LETI, a major French research institution. His research focuses on low-power electronics using heterogeneous devices including modeling, circuit design and control techniques. He has published more than 70 papers in his areas of interest.